\documentclass[aps,prx,reprint,amsmath,amssymb,superscriptaddress,showpacs,nobibnotes]{revtex4-1}

\usepackage{graphicx}
\usepackage{bm}     
\usepackage{hyperref} 
\usepackage{dsfont}    
\usepackage{color}
\usepackage[normalem]{ulem}
\usepackage{cleveref}
\usepackage{multirow} 

\newcommand{\tr}{\text{tr}}
\newcommand{\abs}[1]{\left|{#1}\right|}

\newcommand{\eq}[1]{Eq.~(\ref{#1})}
\newcommand{\fig}[1]{Fig.~\ref{#1}}

\newcommand{\extfig}[1]{Extended Data Fig.~\ref{#1}}
\newcommand{\exttab}[1]{Extended Data Tab.~\ref{#1}}

\def\beq{\begin{eqnarray}}
\def\eeq{\end{eqnarray}}

\newcommand{\expec}[1]{\langle#1\rangle}

\def\S{\mathcal{S}}
\def\R{\mathcal{R}}

\begin{document}

\title{Non-Gaussian quantum states of a multimode light field}

\author{\mbox{Young-Sik Ra}}
\email{youngsikra@gmail.com}
\affiliation{Laboratoire Kastler Brossel, UPMC-Sorbonne Universit\'es, CNRS, ENS-PSL Research University, Coll\`{e}ge de France; 4 place Jussieu, 75252 Paris, France}
\affiliation{Department of Physics, Korea Advanced Institute of Science and Technology (KAIST), Daejeon 34141, Korea}

\author{\mbox{Adrien Dufour}}
\affiliation{Laboratoire Kastler Brossel, UPMC-Sorbonne Universit\'es, CNRS, ENS-PSL Research University, Coll\`{e}ge de France; 4 place Jussieu, 75252 Paris, France}

\author{\mbox{Mattia Walschaers}}
\affiliation{Laboratoire Kastler Brossel, UPMC-Sorbonne Universit\'es, CNRS, ENS-PSL Research University, Coll\`{e}ge de France; 4 place Jussieu, 75252 Paris, France}

\author{\mbox{Cl\'ement Jacquard}}
\affiliation{Laboratoire Kastler Brossel, UPMC-Sorbonne Universit\'es, CNRS, ENS-PSL Research University, Coll\`{e}ge de France; 4 place Jussieu, 75252 Paris, France}

\author{\mbox{Thibault Michel}}
\affiliation{Laboratoire Kastler Brossel, UPMC-Sorbonne Universit\'es, CNRS, ENS-PSL Research University, Coll\`{e}ge de France; 4 place Jussieu, 75252 Paris, France}
\affiliation{Center for Quantum Computation and Communication Technology, Department of Quantum Science,
The Australian National University, Canberra, ACT 0200, Australia}

\author{\mbox{Claude Fabre}}
\affiliation{Laboratoire Kastler Brossel, UPMC-Sorbonne Universit\'es, CNRS, ENS-PSL Research University, Coll\`{e}ge de France; 4 place Jussieu, 75252 Paris, France}

\author{\mbox{Nicolas Treps}}
\affiliation{Laboratoire Kastler Brossel, UPMC-Sorbonne Universit\'es, CNRS, ENS-PSL Research University, Coll\`{e}ge de France; 4 place Jussieu, 75252 Paris, France}

\date{\today}

\begin{abstract} 

Even though Gaussian quantum states of multimode light are promising quantum resources due to their scalability, non-Gaussianity is indispensable for quantum technologies, in particular to reach quantum computational advantage. However, embodying non-Gaussianity in a multimode Gaussian state remains a challenge as controllable non-Gaussian operations are hard to implement in a multimode scenario. Here, we report the first generation of non-Gaussian quantum states of a multimode light field by subtracting a photon in a desired mode from multimode Gaussian states, and observe negativity of the Wigner function. For entangled Gaussian states, we observe that photon subtraction makes non-Gaussianity spread among various entangled modes. In addition to applications in quantum technologies, our results shed new light on non-Gaussian multimode entanglement with particular emphasis on quantum networks.

\end{abstract}

\maketitle







\textbf{
Advanced quantum technologies require scalable and controllable quantum resources~\cite{Andersen:2015dp,Biamonte:2017ic}. Gaussian states of multimode light such as squeezed states and cluster states are scalable quantum systems~\cite{Yokoyama:2013jp,Roslund:2014cb,Chen:2014jx}, which can be generated on demand. However, non-Gaussian features are indispensable in many quantum protocols, especially to reach a quantum computation advantage~\cite{Mari:2012ep}. Embodying non-Gaussianity in a multimode quantum state remains a challenge as non-Gaussian operations generally cannot maintain coherence among multiple modes~\cite{Averchenko:2016gv}. Here, we generate non-Gaussian quantum states of a multimode light field, and observe negativity of the Wigner function in adjustable modes. For this purpose, starting from the deterministic generation of Gaussian entangled states, we use sum-frequency generation to remove a single photon in a computer-controlled coherent superposition of optical modes. We reveal the induced non-Gaussian features and observe how they spread among the entangled modes, depending on the mode in which the photon is subtracted. The resulting non-Gaussian multimode quantum states will have broad applications for universal quantum computing~\cite{Lloyd:1999vz,Menicucci:2006ir}, entanglement distillation~\cite{Eisert:2002ft}, and a nonlocality test~\cite{Plick:2018wc}.
}


Our starting point is a squeezed vacuum state of light, a basic quantum resource for continuous-variable quantum technologies such as quantum-enhanced sensing~\cite{Aasi:2013jb}, deterministic quantum state teleportation~\cite{Takeda:2013hn}, and measurement-based quantum computing~\cite{Menicucci:2006ir}. Recent technological advances have extended the generation and control of squeezed vacuum from a single mode to multiple modes, which enables a determistic generation of large-scale multipartite entangled states~\cite{Yokoyama:2013jp,Roslund:2014cb,Chen:2014jx}. However, such quantum states are intrinsically Gaussian states, which always exhibit Gaussian statistics in electric field quadrature measurements. These states have limitations on quantum applications, e.g., universal quantum computing~\cite{Lloyd:1999vz,Menicucci:2006ir} and entanglement distillation~\cite{Eisert:2002ft}. In particular, the ability to produce a non-Gaussian quantum state is essential to reach quantum advantages~\cite{Mari:2012ep}, which is connected to exotic quantum features of non-Gaussian quantum states. The most profound example thereof is contextuality, which goes hand in hand with negative values of the Wigner function~\cite{Spekkens:2008kc}. Another example is multimode entanglement, which can have conceptually different properties in non-Gaussian states as compared with their Gaussian counterparts~\cite{Valido:2014iv}.

The hybrid approach, which combines continuous-variable and discrete-variable quantum information processing, provides a solution~\cite{Andersen:2015dp}. Subtracting/adding a discrete number of photons~\cite{Wenger:2004cw} or coupling with a discrete-level quantum system~\cite{Vlastakis:2013vi} can generate non-Gaussian states such as a local or non-local superposition of coherent states~\cite{Ourjoumtsev:2006jn,Sychev:2017fq,Ourjoumtsev:2009jh} and hybrid entanglement~\cite{Jeong:2014bl,Morin:2014ip}. Hitherto, this approach has only been successfully applied to a single- or two-mode quantum state, and the extension to highly multimode quantum states remains challenging due to the arduous task of maintaining coherence among multiple modes. For example, the conventional method of photon subtraction for a single-mode quantum state is based on a simple beam splitter~\cite{Wenger:2004cw}; when applied to a multimode quantum state, however, the method results in the generation of a mixed quantum state~\cite{Averchenko:2016gv}.

\begin{figure*}[t]
\centerline{\includegraphics[width=0.8\textwidth]{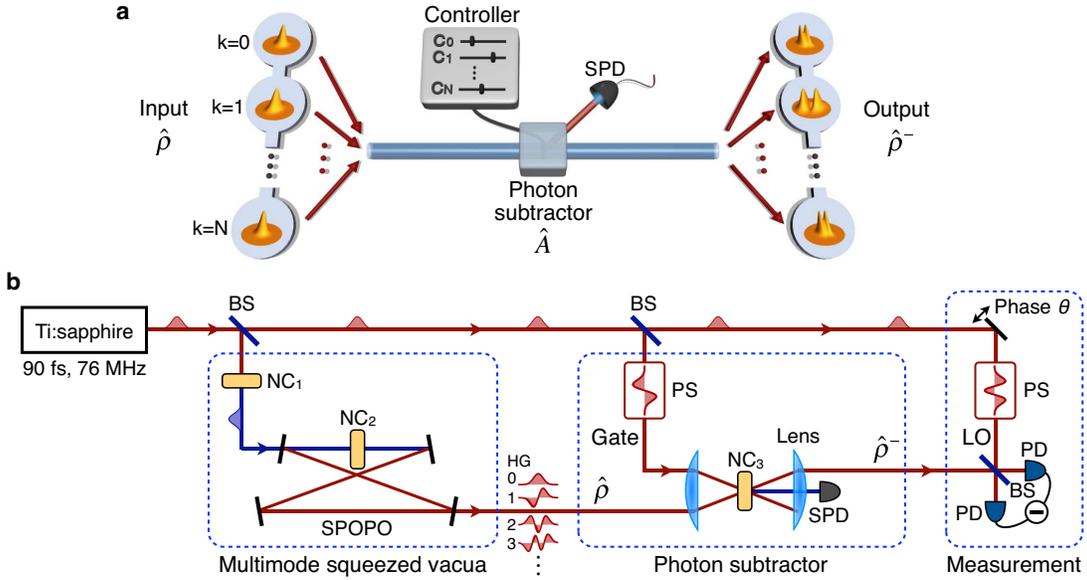}}
\caption{\textbf{Mode-selective photon subtraction from a multimode quantum state.} (a) Concept. Input is a beam containing a Gaussian multimode quantum state $\hat{\rho}$, which can be, in general, a multipartite entangled state. From the input state, we subtract a photon in a specific mode or in a coherent superposition of multiple modes by controlling the complex coefficient $c_k$ for each mode $k$. This process, described by an annihilation operator $\hat{A}$ in \eq{eq:subtraction}, is heralded by a registration of a photon at the single-photon detector (SPD). As a result, the output beam contains a non-Gaussian multimode quantum state $\hat{\rho}^-$. The inset inside each circle is the Wigner function of the reduced quantum state in the associated mode.
(b) Experimental setup. A Ti:sapphire laser produces a beam made of a train of femtosecond pulses, which splits into three beams. One beam is up-converted via second harmonic generation in a second-order nonlinear crystal (NC$_1$), and then, pumps NC$_2$ for a parametric down-conversion process. Synchronously pumped optical parametric oscillator (SPOPO) amplifies the process, which generates twelve-mode squeezed vacua in well-defined time-frequency modes. Another beam is used as a gate for the photon subtractor, and its time-frequency mode is engineered by a pulse shaper (PS). Inside NC$_3$, sum-frequency-interaction between the gate and the multimode squeezed vacua generates an up-converted beam, which is detected by SPD. A photon registration in SPD heralds photon subtraction from the multimode squeezed vacua. The resulting multimode quantum state is measured by homodyne detection with a time-frequency-engineered local oscillator (LO) using another PS. PD: photo diode; BS: beam splitter; $\theta$: phase of LO.
}\label{fig:description}
\end{figure*}

To exploit the full potential of the large-scale entangled states available in the continuous-variable quantum information processing~\cite{Yokoyama:2013jp,Roslund:2014cb,Chen:2014jx}, it is essential that the hybrid approach is made compatible with multimode quantum states, e.g., via photon subtraction operating in multiple modes coherently. The concept of our experiment is illustrated in \fig{fig:description}(a). If we call $\hat\rho$ the density operator of an input multimode quantum state, the output state $\hat{\rho}^{-}$ by a photon subtraction operator $\hat{A}$ becomes
\beq \label{eq:subtraction}
\quad \ \hat\rho^{-}\propto \hat A \hat\rho \hat A^{\dagger},\textrm{\,\,\, where\,\,} \hat{A}=\sum_{k=0} c_k \hat{a}_k.
\eeq
$c_k$ are complex numbers normalized as $\sum \abs{c_k}^2 = 1$, and $\hat{a}_k$ is the annihilation operator for mode $k$. 
Note that $\hat{A}$ is, in general, a coherent superposition of annihilation operators in multiple modes. The ability to experimentally control both the $c_k$ coefficient and the multimode resource $\hat\rho$ is the key to tailor non-Gaussian multimode states and to achieve non-Gaussian entanglement for building non-Gaussian quantum networks~\cite{Walschaers:2017bx,Walschaers:2018wl}.

In our experiment, the controlled generation of non-Gaussian multimode quantum states is performed using quantum frequency combs as a resource. Figure \ref{fig:description}(b) shows the experimental setup, whose details are presented in Methods. The optical modes in which we implement \eq{eq:subtraction} are time-frequency modes~\cite{Ansari:2018uj}. The interest of these modes is that they are co-propagating in the same transverse mode, allowing for a large multimode quantum resource to keep its coherence and to access arbitrary superpositions of modes through elaborate techniques in ultrafast optics. We populate these modes with a highly multimode Gaussian state through a parametric down conversion process~\cite{Roslund:2014cb}. Tailoring the measurement mode basis allows for the generation of versatile multipartite entangled states~\cite{Cai:2017cp}. We combine this resource with a time-frequency mode-dependent photon subtractor to de-Gaussify such multimode Gaussian states.

\begin{figure*}[t]
\centerline{\includegraphics[width=0.99\textwidth]{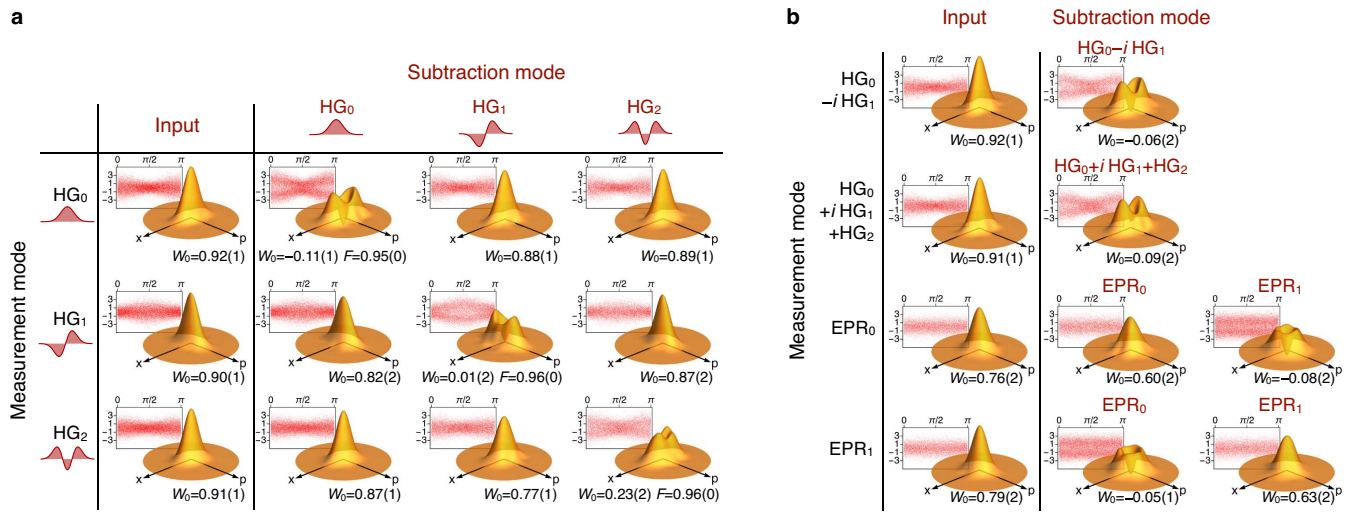}}
\caption{\textbf{Wigner function reconstructed from experimental data.} For a multimode quantum state, photon subtraction and measurement are conducted in (a) HG modes or in (b) superpositions of HG modes. The Wigner function of each mode is represented in the phase space of $x$ and $p$ axes, which are associated with quadrature operators $\hat{x}=\hat{a}+\hat{a}^\dagger$ and $\hat{p}=(\hat{a}-\hat{a}^\dagger)/i$, respectively. The inset behind each Wigner function shows the experimentally obtained quadrature outcomes, where the horizontal and vertical axes represent the phase of local oscillator and a quadrature outcome, respectively; the quadrature outcome of one corresponds to the variance of the vacuum fluctuation. No correction of optical losses is made; for the results by optical loss correction, see \extfig{extfig:datalosscorr}.
 $W_0=2\pi W(0,0)$ is the value of a normalized Wigner function at the origin, and $F$ is the fidelity between an experimental Wigner function and the Wigner function by the ideal photon subtraction to the input state of the corresponding mode. Purities of the Wigner functions in (a), compared with the cases of the ideal photon subtraction, are presented in \exttab{exttab:purity}. For measurements in modes EPR$_0$ and EPR$_1$, the phase of a quadrature outcome is randomized since the associated quantum state is phase insensitive. Errors noted in parentheses are 1 s.~d.
}\label{fig:data}
\end{figure*}

More specifically, our multimode Gaussian resource $\hat\rho$ is a set of independent squeezed vacua whose eigen modes are conveniently approximated by Hermite-Gaussian modes HG$_k$. In order to implement the concepts of \eq{eq:subtraction}, we associate these modes with the annihilation operators $\hat a_k$. Hence, it remains to control the $c_k$ coefficients for the photon subtraction. As shown in \fig{fig:description}(b), this is implemented through a mode-selective sum-frequency generation between the Gaussian resource $\hat{\rho}$ and a gate beam~\cite{Ra:2017ia}. The non-linear interaction is designed such that pulse-shaping the gate allows for the control of the mode of photon subtraction~\cite{Ansari:2018uj}. Finally, detection of a single photon in the up-converted beam heralds the subtraction of a photon in the desired mode from the Gaussian resource.

In practice, in order to implement the operator $\hat A$, the mode of the gate should be set as $v_g = \sum_k (-1)^k c_k~ $HG$_k^{}$~\cite{Ra:2017ia}, which is efficiently performed using a computer controlled pulse shaper. The intensity of the gate governs the efficiency of the operation, and hence the heralding probability. To characterize the generated non-Gaussian multimode quantum state, we employ a homodyne detection that can control the mode of measurement by pulse-shaping the local oscillator.

We first measure the input multimode squeezed vacua without photon subtraction (i.e., no gate field is applied). As expected, the measured state exhibits Gaussian distribution: see the input Wigner functions shown in the first column of~\fig{fig:data}(a). On the other hand, when a single photon is subtracted in HG$_0$ (the second column), we observe that the Wigner function in HG$_0$ becomes non-Gaussian while the Wigner functions in the other modes remain Gaussian. This result shows the mode-selective operation of the photon subtractor necessary for multimode quantum states. The non-Gaussian Wigner function in HG$_0$ exhibits a negative value at the origin ($W_0$), as is required to achieve quantum advantages~\cite{Mari:2012ep}, and the negativity indicates a negative value in the entire multimode Wigner function (see Methods). When a photon is subtracted in HG$_1$ (HG$_2$), we similarly observe a non-Gaussian Wigner function only in HG$_1$ (HG$_2$). Compared with the photon subtraction in HG$_0$, photon subtraction in the higher order modes results in a less non-Gaussian Wigner function. This is mainly because the input state in a higher order mode has a larger optical loss than HG$_0$: we have found a high fidelity ($F$) between a non-Gaussian Wigner function obtained in experiment with the ideal Wigner function calculated by subtracting a photon from the corresponding input state.

Furthermore, the versatility of the experimental setup allows for the computer controlled subtraction of a photon in an arbitrary superposition of modes from a multimode quantum state. As an example, we subtract a photon in a superposition of HG$_0$ and HG$_1$ modes, HG$_0 - i$HG$_1$~\footnote{The normalization constant is omitted for simplicity, and $-i$ is introduced to rotate the $x$-squeezed vacuum in HG$_1$ to $p$-squeezed vacuum such that both HG$_0$ and HG$_1$ have a $p$-squeezed vacuum.}. We now observe a non-Gaussian Wigner function in this mode (first row of \fig{fig:data}(b)). On the other hand, in the orthogonal mode HG$_0 + i$HG$_1$, a Gaussian Wigner function is obtained, and in a partially overlapping mode $i$HG$_1$, an intermediate situation is obtained as expected, see \extfig{extfig:modematching}. When we subtract a photon in a superposition of three modes HG$_0+i$HG$_1$+HG$_2$, we similarly observe a non-Gaussian Wigner function in the same superposed modes (second row of \fig{fig:data}(b)).

This flexibility of the setup allows us to extend photon subtraction to entangled input states. We first investigate an Einstein-Podolsky-Rosen (EPR) entangled state, which exhibits quantum correlations between two superposed modes: EPR$_0=$HG$_0$+HG$_1$ and EPR$_1=$HG$_0-$HG$_1$ (see Methods). The last two rows in \fig{fig:data}(b) show the experimentally obtained Wigner functions. Without photon subtraction, the reduced quantum state in each of EPR$_0$ and EPR$_1$ is a thermal state as expected. When a photon is subtracted in EPR$_0$, the introduced non-Gaussian characteristic appears in the other mode (EPR$_1$) with almost no effect on the mode in which the photon is actually subtracted, and vice versa. In striking contrast with the aforementioned separable input state, the effect of photon subtraction on an entangled state is not localized but is transferred to another mode.

\begin{figure}[t]
\centerline{\includegraphics[width=0.4\textwidth]{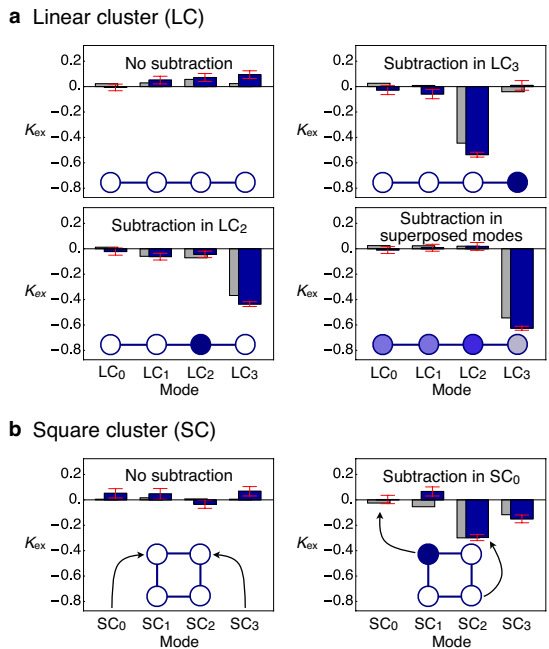}}
\caption{\textbf{Effect of photon subtraction on cluster states.} (a) A linear and (b) a square cluster state. $K_{\rm{ex}}$, defined in \eq{eq:kurtosis}, is the excess kurtosis of quadrature outcomes by randomizing the phase information, where $K_{\rm{ex}}<0$ indicates a non-Gaussian quantum state. For the photon subtraction in the superposed modes (bottom right in (a)), we use $-0.4i \textrm{LC}_0-0.4 \textrm{LC}_1+0.8i \textrm{LC}_2+0.2 \textrm{LC}_3$.
Blue and gray bars are experimental and theoretical values, respectively. The theoretical model is described in Methods. Error bars for experimental data are 1 s. d.
}\label{fig:cluster}
\end{figure}

To open genuine perspectives for applications in quantum technologies, we show the scalability of our approach to larger multimode entangled states. We consider a four-mode linear cluster (LC) state and a four-mode square cluster (SC) state (see Methods), where we denote the four modes of the linear one as LC$_k$ ($k=0,1,2,3$) and the square one as SC$_k$ ($k=0,1,2,3$). Due to the large dimension of the state, full tomography becomes impractical. Thus, we quantify the non-Gaussianity in each mode by evaluating the phase-averaged excess kurtosis (to compare with the kurtosis of 3 by a Gaussian distribution). This is obtained from the quadrature measurement outcomes  $x_1,x_2,..., x_S$ where the measurement phase is randomized:
\beq \label{eq:kurtosis}
K_{\rm{ex}} = \frac{ \frac{1}{S} \sum_{s=1}^{S} x_s^4 } { {(\frac{1}{S} \sum_{s=1}^{S} x_s^2 )}^2 } -3
\eeq
For a quantum state with zero mean ($\expec{\hat{a}}=\expec{\hat{a}^\dagger}=0$), which corresponds to all the quantum states in our experiment, a Gaussian state always exhibits $K_{\rm{ex}}\ge0$, thus $K_{\rm{ex}}<0$ indicates a non-Gaussian state. Figure~\ref{fig:cluster} shows excess kurtosis of the generated states. For the linear cluster state, excess kurtosis in each mode is initially close to zero. When a photon is subtracted in LC$_3$, it is LC$_2$ in which excess kurtosis  becomes highly negative (i.e. non Gaussian) while those in the other modes remain close to 0. For photon subtraction in LC$_2$, only LC$_3$ exhibits a significant negativity. We can concentrate more non-Gaussianity in LC$_3$ by subtracting a photon in a superposition of four modes, as shown in the last figure of~\fig{fig:cluster}(a). For the square cluster state, excess kurtosis without photon subtraction is close to zero in each mode. Photon subtraction in SC$_0$, however, does not affect the distributions in its nearby modes (SC$_1$ and SC$_3$) much, but it introduces non-Gaussianity mostly in SC$_2$ which is two steps away from the mode of photon subtraction.


We have generated non-Gaussian quantum states of a multimode light by subtracting a photon from multimode Gaussian states. The selectivity and the controllability of the mode(s) for the photon subtraction make it possible to extend the non-Gaussianity of a quantum state to the multimode regime, which has been a main obstacle for scalable quantum information processing~\cite{Andersen:2015dp,Biamonte:2017ic}. The availability of non-Gaussian multimode states will stimulate fundamental studies on multipartite entanglement~\cite{Valido:2014iv} and multimode quantumness~\cite{Hudson:1974eh} by going beyond the Gaussian realm, as well as applications in quantum computing~\cite{Lloyd:1999vz,Menicucci:2006ir} and quantum communication~\cite{Eisert:2002ft,Plick:2018wc}. In particular, the observed nontrivial interplay between photon subtraction and cluster states, confirming recent theoretical predictions~\cite{Walschaers:2018wl}, provides new insights into the fields of quantum networks~\cite{Kimble:2008if} and quantum transport~\cite{Walschaers:2016bm}.

\begin{acknowledgements}
This work is supported by the French National Research Agency projects COMB and SPOCQ, the European Union Grant QCUMbER (no. 665148). C.F. and N.T. are members of the Institut Universitaire de France. Y.-S.R. acknowledges support from the European Commission through Marie Sk\l{}odowska-Curie actions (no. 708201) and support from Basic Science Research Program through the National Research Foundation of Korea (NRF) funded by the Ministry of Education (2018R1A6A3A03012129). M.W. acknowledges funding through research fellowship WA 3969/2-1 from the German Research Foundation (DFG).
\end{acknowledgements}



\clearpage

\section{Methods}

\textbf{Experimental details.}
Non-Gaussian multimode quantum states are generated by several nonlinear interactions on femtosecond pulses, as described in~\fig{fig:description}. The fundamental light source is a Ti:Sapphire laser, which produces a train of pulses (duration: 90 fs, central wavelength: 795 nm) at a repetition rate of 76 MHz. The laser beam is split into three beams: one is used for generating a multimode Gaussian state, another for photon subtraction, and the third for the homodyne detection.

The first beam is up-converted to a femtosecond pulse having 397.5-nm central wavelength in NC$_1$ (0.2-mm-thick BiB$_3$O$_6$) which is used as a pump for a parametric-down-conversion process in NC$_2$ (2-mm-thick BiB$_3$O$_6$) inside a cavity,  the SPOPO. The length of the SPOPO is locked to the length of the Ti-Sapphire laser via the Pound-Drever-Hall method, such that the train of pump pulses is synchronized with the down-converted  pulses which circulate inside the SPOPO. Transmittance of the output coupler of the SPOPO is 50\%. The light coming out through the output coupler is a multimode Gaussian state, containing roughly twelve squeezed vacua in orthogonal time-frequency modes~\cite{Roslund:2014cb, Cai:2017cp}. Among the twelve modes, we focus on the first four dominating modes, whose covariance matrix is given in \extfig{extfig:cov}.

For photon subtraction, we perform sum-frequency interaction between the multimode Gaussian state and the second beam from the Ti-Sapphire laser (the gate; 1 mW power) inside NC$_3$ (2.5-mm-thick BiB$_3$O$_6$). Detection of a single photon generated by the sum-frequency interaction heralds photon subtraction from the multimode squeezed vacua, where the time-frequency mode of the gate determines the photon subtraction mode~\cite{Ra:2017ia}. To engineer the time-frequency mode of the gate, we employ a homemade pulse shaper whose core element is a spatial light modulator, having a spectral resolution of 0.2 nm. Conversion efficiency of the nonlinear interaction is 0.1~\%, and we have typically 110 Hz of heralding rate with background noise of 6 Hz.

The last beam is used as the local oscillator (LO) of the homodyne detection to measure the generated quantum state. The measurement mode is the mode of the LO, which is engineered by another pulse shaper having a spectral resolution of 0.2 nm. For each event of photon subtraction, photocurrent difference between the two PDs is sampled every 2 ns during a 2-$\mu s$ time window, and one quadrature outcome is obtained by calculating the dot product between the samples and the double-sided-decaying-shape temporal mode of the SPOPO~~\cite{NeergaardNielsen:2006hl}. To reconstruct a Wigner function in~\fig{fig:data}, we collect 20,000 $\sim$ 30,000 quadrature outcomes. In the case of no photon subtraction, we monitor the variance of the quadrature outcomes. Phase dependance of the quadrature squeezing of the multimode Gaussian state  provides the phase information of the LO relative to this multimode light.

\textbf{Preparation of entangled states.}
We prepare an entangled state by choosing a specific basis of modes in which quantum correlations among desired modes emerge~\cite{Roslund:2014cb, Cai:2017cp}.  In the HG mode basis, even-order (odd-order) modes exhibit $p$-quadrature ($x$-quadrature) squeezed vacuum. To prepare an EPR entangled state, we use a basis of $\textrm{EPR}_0 = \frac{1}{\sqrt{2}}\left(\textrm{HG}_0+\textrm{HG}_1\right)$ and $\textrm{EPR}_1 = \frac{1}{\sqrt{2}}\left(\textrm{HG}_0-\textrm{HG}_1\right)$. We have obtained $\expec{\Delta^2 (\hat{x}^\textrm{EPR}_0 - \hat{x}^\textrm{EPR}_1)} +\expec{\Delta^2(\hat{p}^\textrm{EPR}_0 + \hat{p}^\textrm{EPR}_1)} = 2.51(6) < 4$ (the Duan entanglement criterion~\cite{Duan:2000fw,Simon:2000fd}) and $\expec{\Delta^2 \hat{x}^\textrm{EPR}_{1|0}} \expec{\Delta^2 \hat{p}^\textrm{EPR}_{1|0}}  = 0.71(4)  < 1 $ (the EPR criterion~\cite{Bowen:2003kk}). To prepare a linear cluster state, we use a basis of LC$_k$ ($k=0,1,2,3$) which is obtained by applying a unitary matrix $U^{\textrm{(LC)}}$ to the basis of HG$_k$ ($k=0,1,2,3$), where
\[U^{\textrm{(LC)}}=\left( {\begin{array}{*{20}{r}}
{- 0.344i}&{- 0.421i}&{0.531i}&{0.650i}\\
{0.344}&{ - 0.765}&{ - 0.531}&{0.119}\\
{- 0.765i}&{ - 0.344i}&{- 0.119i}&{- 0.531i}\\
{0.421}&{ - 0.344}&{0.650}&{ - 0.531}
\end{array}} \right).\]
To check correlations among different modes, we use the four nullifiers associated with a linear cluster state~\cite{vanLoock:2007ky,Cai:2017cp}, $\hat{\delta}_k^{\textrm{(LC)}} = \hat{x}_k - \sum_l V_{kl} \hat{p}_l $ ($V_{kl}$ is the adjacency matrix defining the topology of a cluster state, where $V_{kl}=1$ if $k$ and $l$ are connected, and 0 otherwise). They all exhibit a variance less than the vacuum fluctuation: $\expec{ \Delta^2 \hat{\delta}_k^\textrm{{(LC)}}} / \expec{\Delta^2 \hat{\delta}_k^\textrm{(LC)}}_{\textrm{vacuum}} =$ 0.75(2), 0.67(2), 0.68(2), and 0.64(2) for $k =$ 0, 1, 2, and 3, respectively. Similarly, we prepare a square cluster state in the basis of SC$_k$ ($k=0,1,2,3$), which is obtained by applying a unitary matrix $U^{\textrm{(SC)}}$ to the HG basis, where
\[U^{\textrm{(SC)}}=\left( {\begin{array}{*{20}{r}}
{ - 0.316}&{0.632}&{0.707}&{0.000}\\
{0.632i}&{\,0.316i}&{0.000}&{ - 0.707i}\\
{ - 0.316}&{0.632}&{ - 0.707}&{0.000}\\
{0.632i}&{0.316i}&{0.000}&{0.707i}
\end{array}} \right).\]
Each of the four nullifiers $\hat{\delta}_k^{\textrm{(SC)}}$ associated with a square cluster state exhibits a variance less than the vacuum fluctuation: $\expec{\Delta^2 \hat{\delta}_k^\textrm{{(SC)}}} / \expec{\Delta^2 \hat{\delta}_k^\textrm{(SC)}}_{\textrm{vacuum}} =$ 0.72(2), 0.77(2), 0.61(2), and 0.75(2) for $k =$ 0, 1, 2, and 3, respectively.

\textbf{Theoretical model.} To calculate $\hat{\rho}^-$, we model a single-photon subtractor that takes into account experimental imperfections~\cite{Ra:2017ia}:
\beq
\hat{\rho}^- = \R[\hat{\rho}]/\tr{[\R[\hat{\rho}]]};\,\,\,
\R[\hat{\rho}]=w_0\hat{\rho} + (1-w_0) \S[\hat{\rho}],
\label{eq:realsubtractor}
\eeq
where $w_0$ is the weight of background noise (e.g. dark counts) of the SPD which does not alter the input state. $\S[\hat{\rho}]$ is the actual photon subtraction from an input state:
\beq
\S[\hat{\rho}] = \frac{N p_0 -1}{N-1} \, \hat{A} \hat{\rho} \hat{A}^\dagger + \frac{1-p_0}{N-1}\sum_{k=0}^{N-1} \, \hat{a}_k \, \hat{\rho} \, \hat{a}_k^\dagger,
\eeq
where $N$ is the number of modes. $\hat{A}$ is the desired photon subtraction in~\eq{eq:subtraction}, whose weight is $p_0$; the remaining weight, $1-p_0$, corresponds to the photon subtraction in the incoherent mixture of the other modes with equal contribution. To consider the experimental conditions, we use $w_0=0.0094$, $p_0=0.95$, $N=4$, and $\hat{\rho}$ in~\extfig{extfig:cov}. $K_{\textrm{ex}}$ of $\hat{\rho}^-$ in a specific mode is then obtained by following the method presented in Refs.~\cite{Walschaers:2017bx,Walschaers:2018wl}, which is based on calculating multimode correlation functions.

\textbf{Negativity of a multimode Wigner function.}
A negative value in a single-mode Wigner function, which is reduced from a multimode Wigner function, is sufficient to show a negative value in the multimode Wigner function. The proof is straightforward by considering the contraposition of the statement: a non-negative multimode Wigner function always leads to a non-negative Wigner function when reduced to a single mode.
Consider a reduced phase space $\mathbb{R}^2$ of an arbitrary single mode defining quadratures $(x_0,p_0)$ and the entire phase space $\mathbb{R}^{2N}$ of $N$ modes defining quadratures $(x_0,p_0, ..., x_{N-1},p_{N-1})$. The Wigner function in the $N$ modes is assumed to be non negative: $W^{(N)}(x_0,p_0, ..., x_{N-1},p_{N-1})\ge 0$. Then, the Wigner function in the reduced phase space $W(x_0,p_0)$ can be obtained by integrating the multimode Wigner function over the phase space of $(x_1,p_1, ..., x_{N-1},p_{N-1})$:
\beq
 W(x_0,p_0)=\int_{\mathbb{R}^{2(N-1)}} dx_1dp_1...dx_{N-1}dp_{N-1}~~W^{(N)}.\nonumber
\eeq
As $W^{(N)}\ge 0$, the reduced Wigner function $W(x_0,p_0)$ cannot have a negative value.

\clearpage

\setcounter{figure}{0}
\makeatletter 
\renewcommand{\figurename}{Extended Data Fig.}
\renewcommand{\thefigure}{\@arabic\c@figure}
\makeatother

\setcounter{table}{0}
\makeatletter 
\renewcommand{\tablename}{Extended Data Table}
\renewcommand{\thetable}{\@arabic\c@table}
\makeatother

\renewcommand{\arraystretch}{1.0}
\begin{table*}[h]
\begin{tabular}{ccccccccccccc}
&&&\multicolumn{3}{c}{Subtraction mode}
\\
~~&~~&\multicolumn{1}{|c|}{~~Input~~}&~~HG$_0$~~&~~HG$_1$&~~HG$_2$~~
\\\cline{2-6}
\multirow{3}{*}{\shortstack{~Measurement~\\mode}}&\multicolumn{1}{c|}{~~HG$_0$~~}&\multicolumn{1}{c|}{~~0.91(1)~~}&~0.45(0) [0.53(2)]~&0.88(1)&0.87(1)\\
&\multicolumn{1}{c|}{~~HG$_1$~~}&\multicolumn{1}{c|}{~~0.90(1)~~}&0.82(1)&~0.47(0) [0.49(1)]~&0.86(1)\\
&\multicolumn{1}{c|}{~~HG$_2$~~}&\multicolumn{1}{c|}{~~0.91(1)~~}&0.87(1)&0.77(1)&~0.49(0) [0.50(2)]~
\end{tabular}
\caption{
\textbf{Purities of the Wigner functions in \fig{fig:data}(a).} For comparison, the purity of a Wigner function by the ideal photon subtraction is provided in square brackets, which agrees well with the experimental result. Low purity in a photon-subtracted mode is attributed to a non-ideal input state~\cite{Ourjoumtsev:2006jn}. No optical loss is corrected in the calculation. Errors noted in parentheses are 1 s.~d.
}\label{exttab:purity}
\end{table*} 
\renewcommand{\arraystretch}{1}

\begin{figure*}[h]
\centerline{\includegraphics[width=0.99\textwidth]{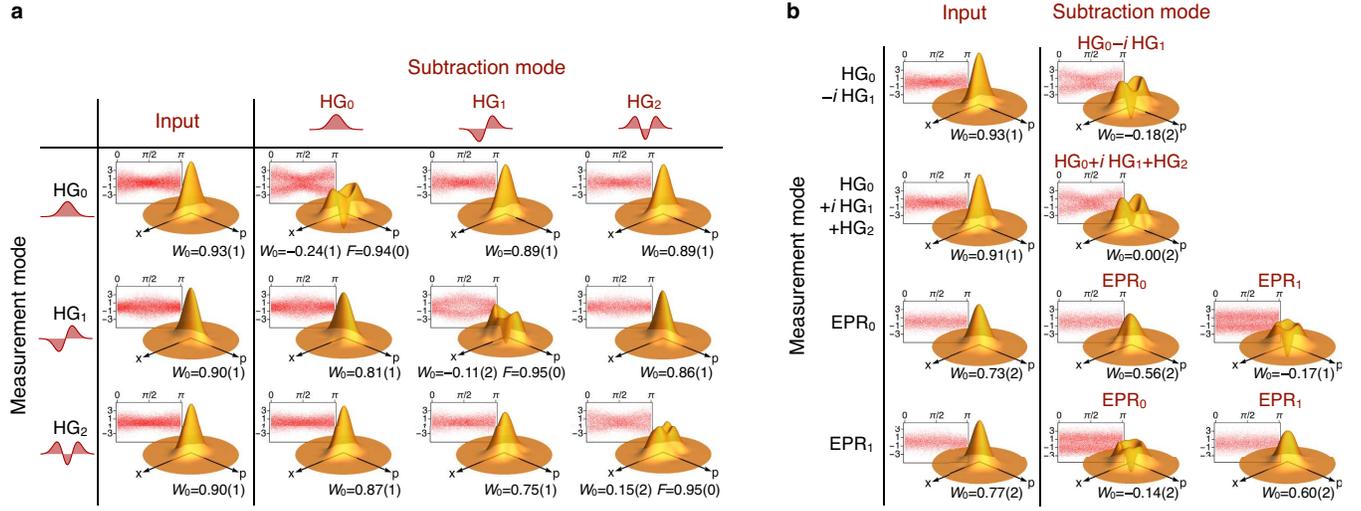}}
\caption{\textbf{Wigner function reconstructed with optical loss correction.} Optical loss in the homodyne detection ($12.5\%$) has been corrected. Comparing with \fig{fig:data}, non-Gaussian Wigner functions show reduced $W_0$. Errors noted in parentheses are 1 s.~d.}\label{extfig:datalosscorr}
\end{figure*}

\begin{figure}[h]
\centerline{\includegraphics[width=0.4\textwidth]{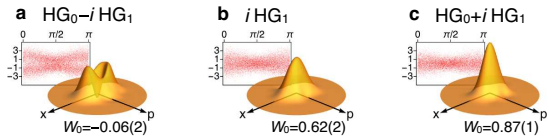}}
\caption{\textbf{Effect of mode mismatch between photon subtraction and measurement.} When a single photon is subtracted in $\textrm{HG}_0 -i\textrm{HG}_1$, a Wigner function (without optical loss correction) is obtained in a measurement mode having (a) full match ($\textrm{HG}_0 -i\textrm{HG}_1$), (b) partial match ($i\textrm{HG}_1$), and (c) no match ($\textrm{HG}_0 +i\textrm{HG}_1$). Errors noted in parentheses are 1 s.~d.
}\label{extfig:modematching}
\end{figure}

\begin{figure}[h]
\centerline{\includegraphics[width=0.4\textwidth]{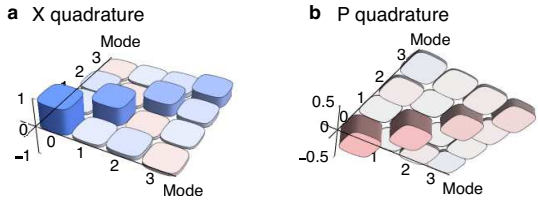}}
\caption{\textbf{Experimental covariance matrix.} (a) is for $x$ quadratures, seen from above, and (b) is for $p$ quadratures, seen from below. Mode indexes are $\textrm{HG}_0$, $i\textrm{HG}_1$, $\textrm{HG}_2$, and $i\textrm{HG}_3$, where $i$ is added for the odd-index HG modes to have $p$-squeezed vacua in all modes. For clarity, the vacuum noise (corresponding to the identity matrix) is subtracted from the covariance matrix. In the covariance matrix, variances of ($x$, $p$) quadratures are ($2.8\textrm{ dB}$, $-1.8\textrm{ dB}$) in mode 0, ($2.1\textrm{ dB}$, $-1.6\textrm{ dB}$) in mode 1, ($1.6\textrm{ dB}$, $-1.0\textrm{ dB}$) in mode 2, and ($1.4\textrm{ dB}$, $-0.7\textrm{ dB}$) in mode 3.
}\label{extfig:cov}
\end{figure}

\end{document}